\documentclass[nofootinbib,twocolumn,showpacs,preprintnumbers,pre,aps]{revtex4-1}

\usepackage{graphicx}
\usepackage{bm}
\usepackage{amsmath}
\usepackage{amssymb}
\usepackage{color}

\begin{document}

\title{Correlated lateral phase separations in stacks of lipid membranes}%

\author{Takuma Hoshino$^{1,2,3}$}\email{hoshino-takuma@ed.tmu.ac.jp}

\author{Shigeyuki Komura$^{1,3}$}\email{komura@tmu.ac.jp}

\author{David Andelman$^{2,3}$}\email{andelman@post.tau.ac.il}

\affiliation{
$^{1}$Department of Chemistry, Graduate School of Science and Engineering,
Tokyo Metropolitan University, Tokyo 192-0397, Japan\\
$^{2}$Raymond and Beverly Sackler School of Physics and Astronomy,
Tel Aviv University, Ramat Aviv, Tel Aviv 69978, Israel\\
$^{3}$Kavli Institute for Theoretical Physics China, CAS,
Beijing 100190, China}

\date{\today}

\begin{abstract}
Motivated by the experimental study of Tayebi \textit{et al.}\ [\textit{Nature Mater.}
\textbf{11}, 1074 (2012)] on phase separation of stacked multi-component lipid bilayers,
we propose a model composed of stacked two-dimensional Ising spins.
We study both its static and dynamical features using Monte Carlo simulations
with Kawasaki spin exchange dynamics that conserves the order parameter.
We show that at thermodynamical equilibrium, due to strong inter-layer correlations,
the system forms a continuous columnar structure for any finite
interaction across adjacent layers. Furthermore,
the phase separation shows a faster dynamics as the inter-layer interaction
is increased. This temporal behavior is mainly due to an effective deeper temperature
quench because of the larger value of the critical temperature, $T_{\rm c}$, for
larger inter-layer interaction.
When the temperature ratio, $T/T_{\rm c}$, is kept fixed, the temporal growth exponent
does not increase and even slightly decreases as function of the increased inter-layer
interaction.
\end{abstract}

\maketitle

\section{Introduction}
\label{sec:introduction}

Biological membranes are constructed out of two monolayers (leaflets) arranged
in a back-to-back configuration.
They are mainly composed of phospholipids but contain also other
molecules such as cholesterol, glyco-sugars, and proteins~\cite{AlbertsBook}.
In living organisms, these membranes can form multi-lamellar stacks known
as lamellar bodies~\cite{Schmitz}.
Examples of such highly folded membranous structures are thylakoid membranes of
photosynthetic cyanobacteria or plant chloroplasts, and stratum corneum of
human skin.
Since multilamellar structures can combine single membrane functions in series,
they offer possibilities for novel applications in photonics and as bio-sensors.

Over the last decade, many studies have been performed on artificial giant
unilamellar vesicles (GUVs) composed of ternary mixtures of saturated lipid
such as sphingomyelin, unsaturated lipid such as DOPC
(1,2-dioleoyl-sn-glycero-3-phosphocholine) and cholesterol~\cite{VK05,SK_DA_Review}.
By decreasing temperature, these ternary mixtures undergo a lateral phase separation,
where a liquid-disordered (L$_{\rm d}$) phase coexists with a liquid-ordered (L$_{\rm o}$) one.
It is known that the L$_{\rm o}$ phase is rich in saturated lipid and cholesterol,
while the L$_{\rm d}$ phase is rich in the unsaturated lipid.

In a recent experimental study, Tayebi \textit{et al.}~\cite{Tayebi12} reported
that a stack (typically composed of several hundred layers) of multicomponent
lipid bilayers with phase-separated domains exhibits inter-layer columnar ordering.
Using ternary mixtures of sphingomyelin, DOPC and cholesterol, it was observed
that domains in stacked bilayers align one on top of the other, thereby forming
an uninterrupted columnar ordering across hundreds of bilayer membranes.
Such a cooperative multilayer epitaxy was attributed to the interplay between
intra-layer domain growth and inter-layer coupling.
{The formation of columnar structures in stacked bilayers is 
important because it allows for electrical currents and transport processes to 
pass through many transmembrane channels in a cooperative and efficient manner. 
Other possible applications of the columnar ordering can be as templates for 
membrane protein crystallization, which is necessary for X-ray structural 
analysis of membrane proteins incorporated in bilayers.}

As far as the dynamics of phase separation in stacks of membranes is concerned,
the temporal evolution of the average inplane domain size, $R$, was shown to
obey a power-law growth, $R \sim t^{\alpha}$ with
$\alpha \approx 0.455$~\cite{Tayebi12}.
This exponent is larger than the value obtained using GUVs with a single bilayer,
for which the reported
experimental value is $\alpha \approx 0.28 \pm 0.05$~\cite{Stanich}.
{Hence, Tayebi \textit{et al.}\ concluded that the inplane 
domain growth in each of the bilayers of the stack is faster, as compared to the domain growth in GUVs.”}

In a subsequent paper~\cite{Tayebi13}, a model based on regular
solution theory, which  takes into account the inter-lamellar coupling of inplane
phase-separated domains, was proposed.
The calculated phase diagram was presented in terms of intra-layer and inter-layer
coupling parameters, and contains three different regions: (i) a ``one-phase" region
in which the system does not exhibit phase separation; (ii) a ``two-phase" region
in which two phases coexist and domains in different layers along the normal $z$-direction
are completely aligned and have the same composition in the various layers,
and (iii) a ``multi-phase" region in which there are unaligned inplane domains with
\textit{different composition} in the different layers.
According to Ref.~\cite{Tayebi13}, the transition line between the ``two-phase" and
``multi-phase" regions strongly depends on the number of layers in the stack which
was varied up to ten layers.

\begin{figure}[tbh]
\begin{center}
\includegraphics[scale=0.35]{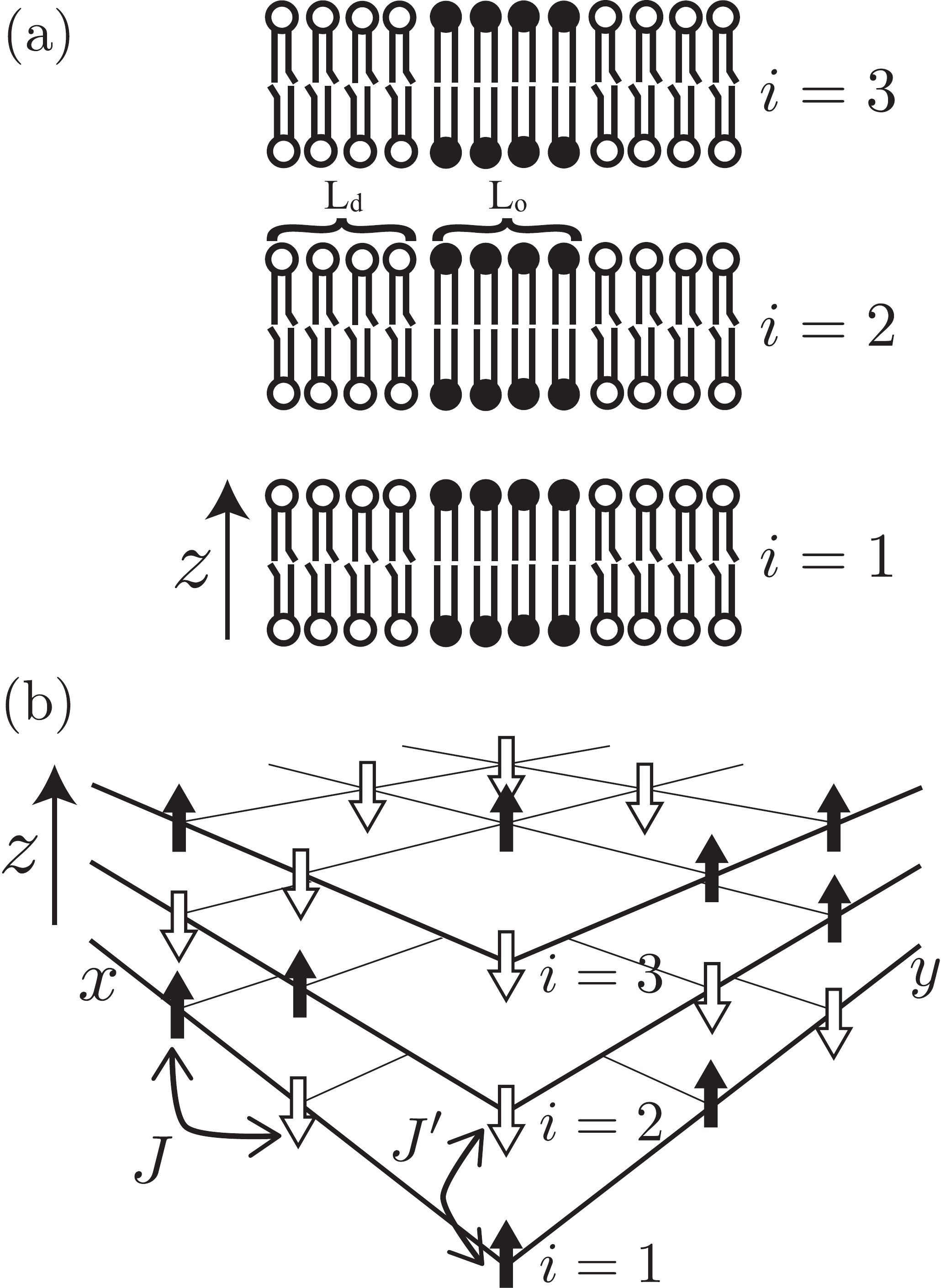}
\end{center}
\caption{\textsf{
(a) Schematic illustration of a stack of binary membranes, taken here as a stack of three
bilayers in the $z$-direction.
Each bilayer is composed of two identical leaflets containing saturated lipids (A, black)
and unsaturated lipids (B, white).
Saturated and unsaturated lipids typically form L$_{\rm o}$ and L$_{\rm d}$
phases, respectively.
As the lipid molecules are not allowed to exchange between different bilayers,
their composition in each bilayer is fixed.
(b) The stacked two-dimensional (2d) Ising model.
Here the bilayer structure of each membrane is neglected.
Lipid A and B correspond to spin up (black) and spin down (white), respectively.
$J$ is the coupling parameter between nearest-neighbor spins in the same layer,
while $J'$ is the coupling parameter between spins belonging to two nearest-neighboring layers.
}}
\label{fig1}
\end{figure}

Being motivated by these works~\cite{Tayebi12,Tayebi13}, we investigate
the correlation between lateral phase separation in a stack of multi-layer
membranes using a spin
model called the \textit{stacked two-dimensional (2d) Ising model.}
{This is the simplest model to describe a stack of binary membranes
composed of two types of lipids.}
The model is the same as the \textit{anisotropic three-dimensional (3d) Ising model}
for a finite stack in the $z$-direction.
The important difference between the two models is that in the former the order
parameter (magnetization) in each layer is conserved.
This requirement is based on the experimental fact that the A/B lipid composition
in each layer almost does not change during experimental times.

In our model, we study the thermodynamical equilibrium features using Monte
Carlo (MC) simulations.
{
The main reason that we performed MC simulations rather than analyzing the 
mean-field free energy describing phase separation (as studied in Ref.~\cite{Tayebi13}), 
is to allow us to investigate the 
role of thermal fluctuations 
on the inter-layer domain correlation in stacked membranes.}
We show that the domains in each layer are correlated along the vertical $z$-direction,
for any finite value of the inter-layer interaction is positive, i.e., $J'>0$.
Hence, the system is either in a one- or two-phase state in equilibrium, and in our
model the ``multi-phase" state is not obtained in the thermodynamic limit of infinite
lateral size, as long as the inter-layer coupling $J'>0$.
As anticipated, it is found that the phase-transition temperature, $T_{\rm c}(J')$,
increases as function of the inter-layer interaction parameter.

We also investigate the dynamics of phase separation at fixed temperature $T$
in the two-phase coexistence region.
We show that the accelerated temporal behavior of the phase separation for the
stack is mainly driven by the increase of the temperature quench,
$\Delta T=T_{\rm c}(J') -T$, because $T_{\rm c}(J')$ becomes larger for larger $J'$.
However, if the ratio $T/T_{\rm c}(J')$ is kept fixed, the dynamics of the
phase separation {is actually slower} for larger values of the
inter-layer coupling, $J'$.

\begin{figure*}[tbh]
\begin{center}
\includegraphics[scale=0.6]{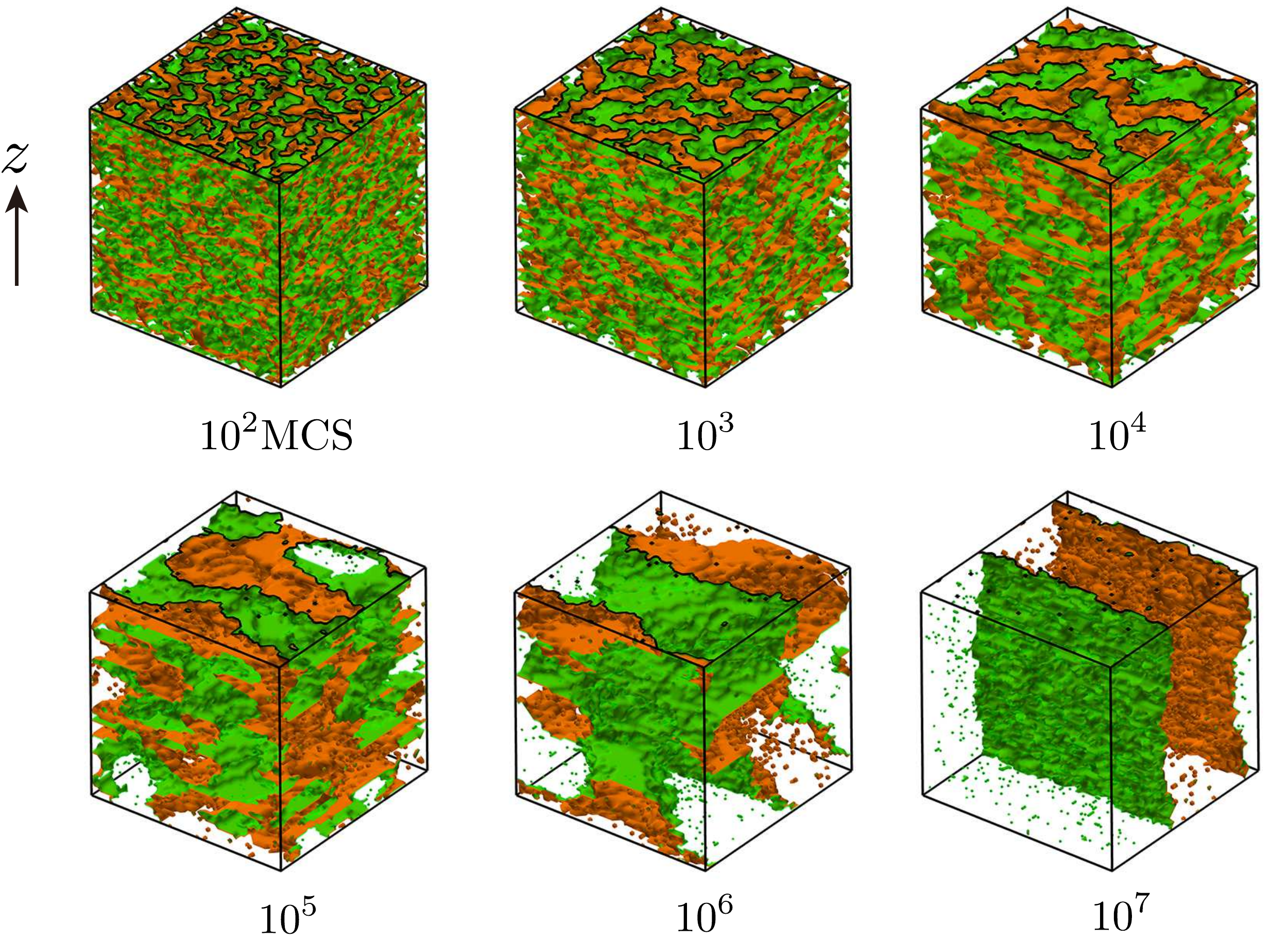}
\end{center}
\caption{\textsf{
Time evolution of phase separated domains in the stacked 2d Ising model at
different MC steps for $\lambda=0.1$ and $T/J=1.63$.
The other parameters are $\overline{S}_{i,\boldsymbol{\rho}}=0$ and $L=L_z=64$.
For presentation purposes, only the interfaces of domain boundaries are shown, and the two different sides
of the interfaces are represented by green and brown.
The system is fully equilibrated after about 10$^7$ MCS.
}}
\label{fig2}
\end{figure*}

In the next section, we describe the stacked 2d Ising model and review the MC
simulation method.
In Sec.~\ref{sec:static}, we  present the equilibrium properties of the
model, and discuss the condition for domain columnar ordering.
Section~\ref{sec:dynamics} describes the dynamics of domain growth for different
values of the inter-layer interaction, and it is compared with a previous theoretical work.

\section{Model and simulation technique}
\label{sec:model}

In our simulations, we use the \textit{stacked 2d Ising model}, shown in
Fig.~\ref{fig1}(a).
We consider a stack of two-component lipid bilayer membranes composed of an A/B
lipid mixture, although the experimental systems often consist of ternary
lipid/cholesterol mixtures.
This simplification does not affect the essential feature of the lateral phase
separation.
Another simplification is that we treat only symmetric bilayers where the
composition of the two leaflets is identical.
Hence, each lipid bilayer having a finite thickness can be mapped into a 2d
Ising model with conserved magnetization, expressing the fact that no lipid is allowed
to exchange across layers.
The 2d Ising layers are stacked in the $z$-direction, and they interact with their
two nearest-neighboring layers, as depicted in Fig.~\ref{fig1}(b).

The Hamiltonian of this stacked and coupled 2d Ising system can be written as:
\begin{align}
H = & -J \sum_{i, \langle \boldsymbol{\rho}, \boldsymbol{\rho}' \rangle}
S_{i,\boldsymbol{\rho}} S_{i,\boldsymbol{\rho}'}
-J' \sum_{i, \boldsymbol{\rho}}
S_{i,\boldsymbol{\rho}} S_{i+1,\boldsymbol{\rho}} \nonumber \\
& - \sum_{i,\boldsymbol{\rho}} \mu_i S_{i,\boldsymbol{\rho}},
\label{hamiltonian}
\end{align}
where up/down values of the spin, $S_{i,\boldsymbol{\rho}}=\pm 1$,
at $\boldsymbol{\rho}=(x,y)$ in the $i$-th layer corresponds
to a lattice site occupied by an A or B lipids, respectively.
The coupling between nearest-neighbor spins in the $xy$-plane
(denoted by $\langle \boldsymbol{\rho}, \boldsymbol{\rho}' \rangle$) is $J$,
while the coupling with the nearest-neighbor spins across layers in the
$z$-direction is $J'$.
{The physical origin of the inter-layer interaction $J'$ is
primarily attributed to direct van der Waals attractive interactions acting
between neighboring bilayers~\cite{Israelachivili}.
Other non-specific interactions, such as electrostatic and/or hydration
interactions, can be taken into account through the second virial coefficient
and will affect the value of $J'$ as well~\cite{MR92,KA04}.}
Throughout this paper, we shall use the dimensionless ratio defined by
$\lambda \equiv J'/J$ as a measure of the inter-layer coupling strength.

In the above Hamiltonian, $\mu_i$ is the external field (chemical potential),
which fixes the average magnetization (A/B composition) in the $i$-th layer.
Although $\mu_i$ can, in general, take different values for different layers,
we consider here the case where all of them are the same, $\mu_i=\mu$, fixing
the same value of lipid composition in all layers.
This assumption holds also for the dynamical states since we do not allow
the lipids to be exchanged across different layers.
{The average order parameter (A/B composition) in the $i$-th
layer is denoted by $\overline{S}_{i,\boldsymbol{\rho}}$, and throughout this
paper (except in Fig.~\ref{fig5}(b)) we choose $\overline{S}_{i,\boldsymbol{\rho}}=0$,
which corresponds to a symmetric $1{:}1$ A/B lipid mixture, \textit{i.e.},
at the critical composition.
This is equivalent to setting the value of the chemical potential to zero, i.e., $\mu=0$.}

The present model is related to the {\it anisotropic 3d Ising model} for a finite slab.
The special case of $\lambda=1$ corresponds to the isotropic 3d Ising model,
whereas for $\lambda=0$ the stack is composed of non-interacting 2d Ising layers.
One interesting issue related to the anisotropic model, $0<\lambda<1$, is the
crossover from 2d to 3d critical behavior~\cite{lee01} that will be explored below.
{We also note that the stacked 2d Ising model has been studied a 
great deal in connection with multilayer adsorption phenomena on attractive 
substrates~\cite{Oliveira78,Pandit82}, but not in the context of layers of binary
mixtures with conserved magnetization (order parameter) as studied in this paper.}

We investigate both the statics and dynamics of a stack of
membranes based on the Hamiltonian presented in Eq.~(\ref{hamiltonian}).
We employ MC simulations for classical Ising spins on a finite
$L \times L \times L_z$ lattice.
Periodic boundary conditions are used in all three directions.
The spin configurations are updated using Kawasaki exchange dynamics~\cite{Kawasaki}
in order to conserve the magnetization in each layer.
This is based on the experimental fact that lipids almost do not exchange across
different layers.
Hence, their A/B inplane composition is fixed during experimental times.

The MC simulations presented here are performed in the following way.
At each MC trial step, a site on the 3d lattice and one of its
nearest neighbors in the {\it same} layer are chosen at random.
If the two spins are alike, a new site is again chosen at random.
This process is repeated until two unlike nearest neighbor spins are found.
Then, the probability of exchanging the two spins is determined by the standard Metropolis
algorithm~\cite{Landau_Binder}.
If the energy difference due to the spin exchange becomes negative, i.e.,
$\Delta E < 0$, we accept the exchange.
Otherwise, we accept the exchange with a probability $\exp(-\Delta E/T)$,
where $T$ is the temperature and the Boltzmann constant, $k_{\rm B}$, was set to unity.

In one Monte Carlo step (MCS), this procedure is repeated $L\times L\times L_z$ times.
The MC simulations are carried out by annealing the temperature gradually from
an initial infinite temperature for which the spin configurations are completely
disordered and uncorrelated.
The first $10^5$ (or in some cases up to $10^6$) MCS are discarded in order to reach thermal equilibration.
Furthermore, to avoid correlations between different equilibrated configurations,
measurements are taken every 20 MCS, and we averaged over 10$^5$ independent system
configurations, in order to obtain sufficient statistics.

In order to investigate the phase separation dynamics, we monitor the domain
coarsening as a function of time (MCS) at a constant temperature below $T_{\rm c}$.
An example of a typical time evolution of phase separation is presented in
Fig.~\ref{fig2} for $\lambda=0.1$, $T/J=1.63$ and $L=L_z=64$, where six snap-shots are shown from $10^2$\,MCS till $10^7$\,MCS.
For clarity purposes, only the  boundaries between domains of spin up (rich in lipid A)
and spin down (rich in lipid B) are shown. In the initial time steps, the phase separation occurs inplane, and the domains coarsen without much out-of-plane coupling (due to the rather small value of $\lambda=0.1$). As time evolves, the inplane coarsening is also followed by out-of-plane columnar ordering, where the lipid A (and lipid B) rich domains are highly correlated along the $z$-direction.
This is clearly seen for the fully equilibrated configuration occurring after about 10$^7$ MCS (last snap-shot). Here the two color boundaries, represent the two sides of the domain boundaries (while the inside of the domain is not shown). The boundaries look like extended interfaces separating
inplane domains that are vertically connected along the $z$-direction, in agreement with experiment~\cite{Tayebi12}.

\section{Static properties of the stacked domains}
\label{sec:static}

\begin{figure}[tbh]
\begin{center}
\includegraphics[scale=0.38]{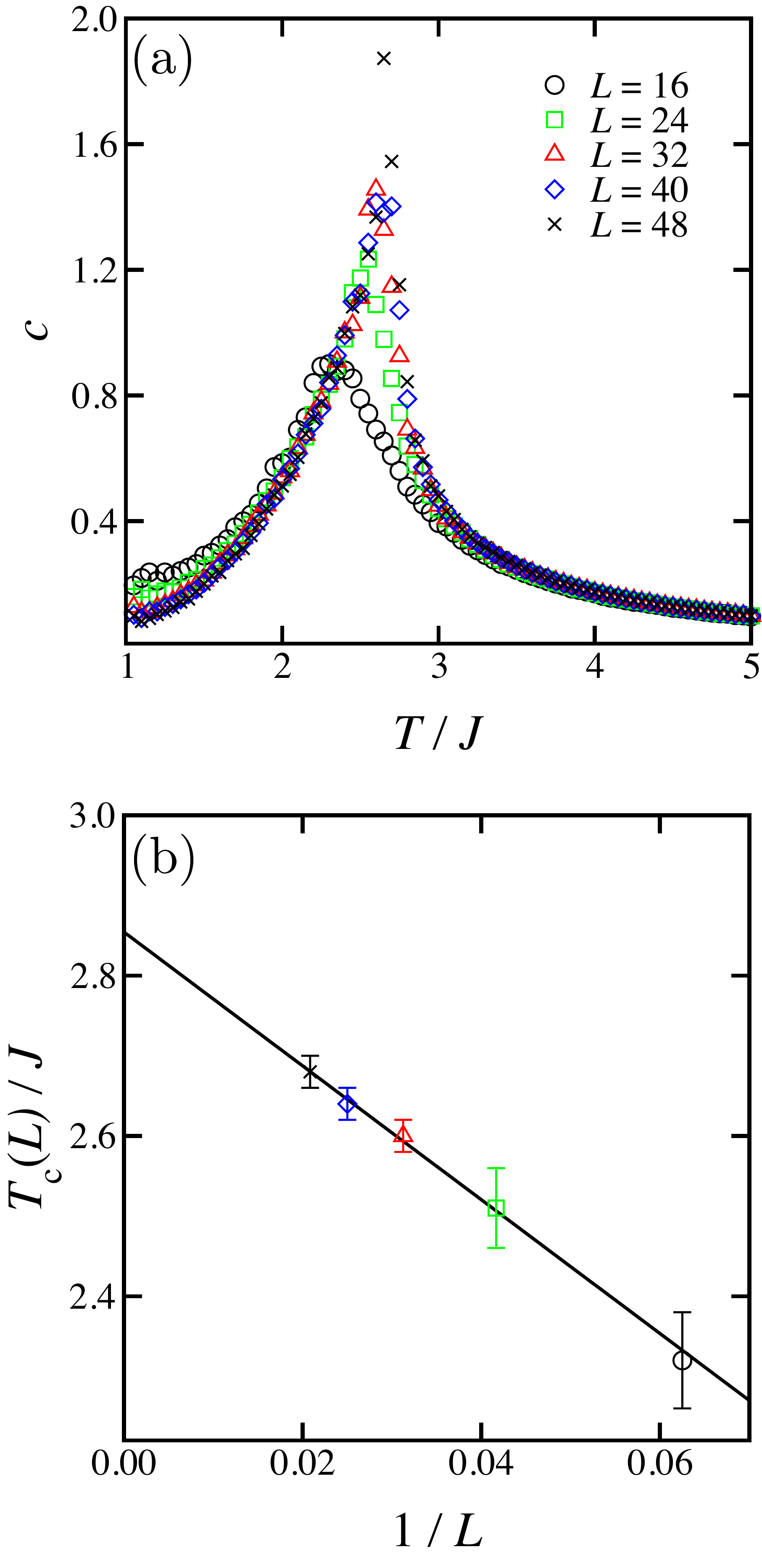}
\end{center}
\caption{\textsf{
(a) Specific heat per lattice site, $c$, as function of the dimensionless temperature
$T/J$, for different lateral system-size $L=16, 24, 32, 40, 48$.
The other parameters are $\overline{S}_{i,\boldsymbol{\rho}}=0$, $\lambda=0.1$
and $L_z=8$.
For each system size, the peak position of $c$ is identified with an effective ``phase transition"
temperature.
(b) Finite-size scaling analysis of the phase-transition temperature, $T_{\rm c}(L)/J$ for $\lambda=0.1$.
The apparent phase-transition temperature is plotted as a function of $1/L$.
The solid line is the fit given by Eq.~(\ref{finite}) with $\nu=1$ (see text).
The extrapolated value for the critical temperature is $T_{\rm c}(\lambda=0.1)/J= 2.85$.
}}
\label{fig3}
\end{figure}

In order to determine the phase-transition temperature and obtain the
corresponding phase diagram, we compute the specific heat per lattice site defined as
\begin{equation}
c= \frac{1}{L^2 L_z} \frac{1}{T^2}
\left( \langle H^2 \rangle - \langle H \rangle^2 \right),
\end{equation}
where $H$ is given by Eq.~(\ref{hamiltonian}) and $\langle \cdots \rangle$ indicates
an ensemble average.
{We note again that the above specific heat is calculated
at constant magnetization (corresponding to constant lipid concentration in our model) of each layer.}
In our simulations, the ensemble average is taken by averaging over
independent equilibrium spin
configurations as explained in Sec.~\ref{sec:model}.
For a given system size and  dimensionless ratio $\lambda$, we calculate $c$
as function of the dimensionless temperature $T/J$.
Such a dependence of $c$ on $T/J$ is presented in Fig.~\ref{fig3}(a) for several
lateral system-sizes, $L$, and for $\lambda=0.1$, $L_z=8$, recalling that $L_z$
is the number of layers of the 3d stack.

\begin{figure*}[tbh]
\begin{center}
\includegraphics[scale=0.3]{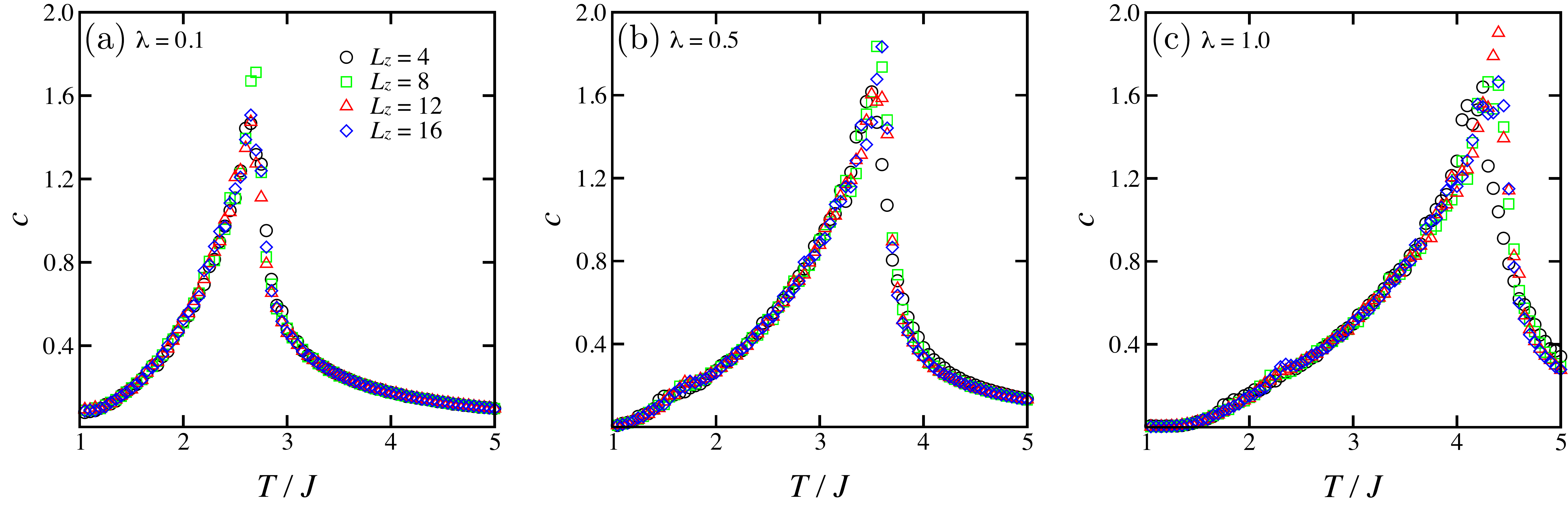}
\end{center}
\caption{\textsf{
Specific heat per lattice site, $c$, as function of the dimensionless temperature
$T/J$, for different systems size $L_z=4, 8, 12, 16$
{for (a) $\lambda=0.1$, (b) $\lambda=0.5$ and (c) $\lambda=1$.}
The other parameters are $\overline{S}_{i,\boldsymbol{\rho}}=0$ and $L=48$.
{The observed peak position:
$T/J \approx 2.65$ in (a), 3.30 in (b) and 4.10 in (c), is almost independent of $L_z$,
at least for $L_z \ge 8$.}
}}
\label{fig4}
\end{figure*}

\begin{figure*}[tbh]
\begin{center}
\includegraphics[scale=0.3]{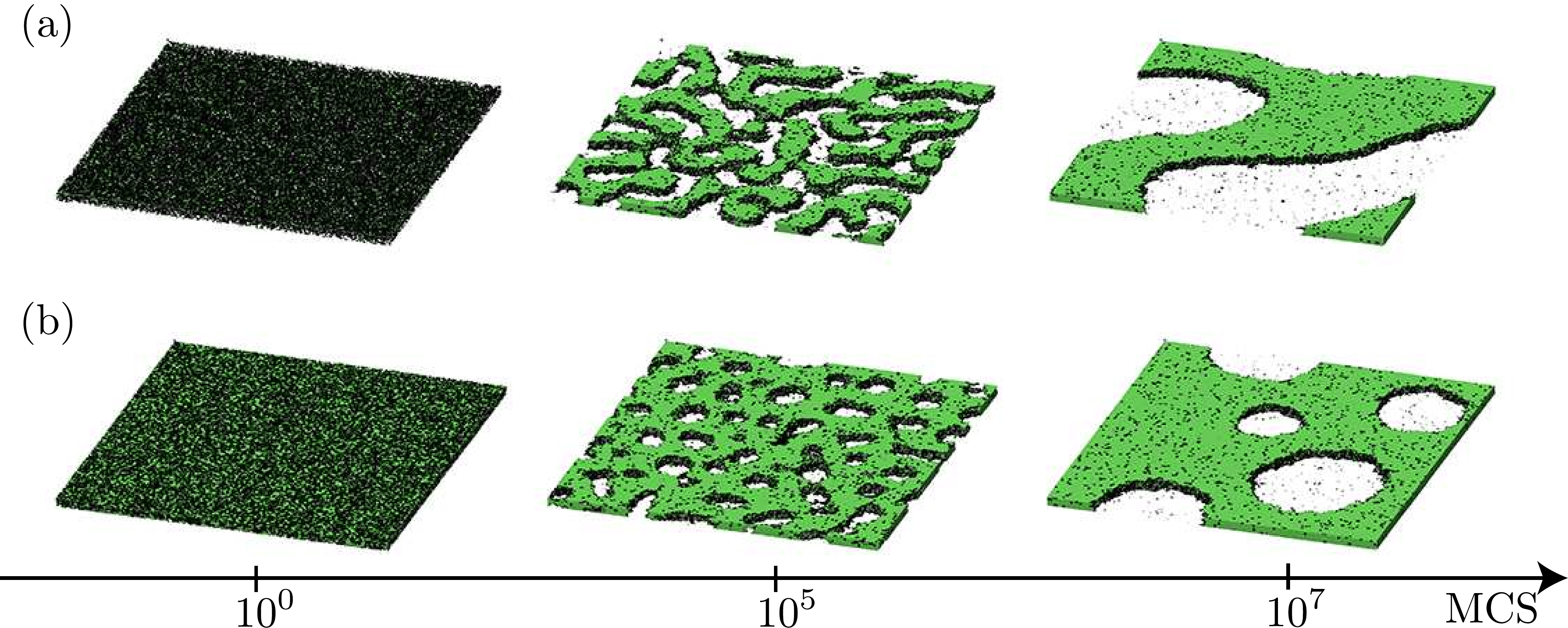}
\end{center}
\caption{\textsf{
{
Time evolution of phase-separated domains in a stacked 2d Ising model of
eight layers, $L_z=8$, at different MC steps for (a) $\overline{S}_{i,\boldsymbol{\rho}}=0$ and
(b) $\overline{S}_{i,\boldsymbol{\rho}}=0.4$.
The other parameters are $\lambda=0.1$, $T/J=2.0$ and $L=256$.}
}}
\label{fig5}
\end{figure*}

\begin{figure}[tbh]
\begin{center}
\includegraphics[scale=0.38]{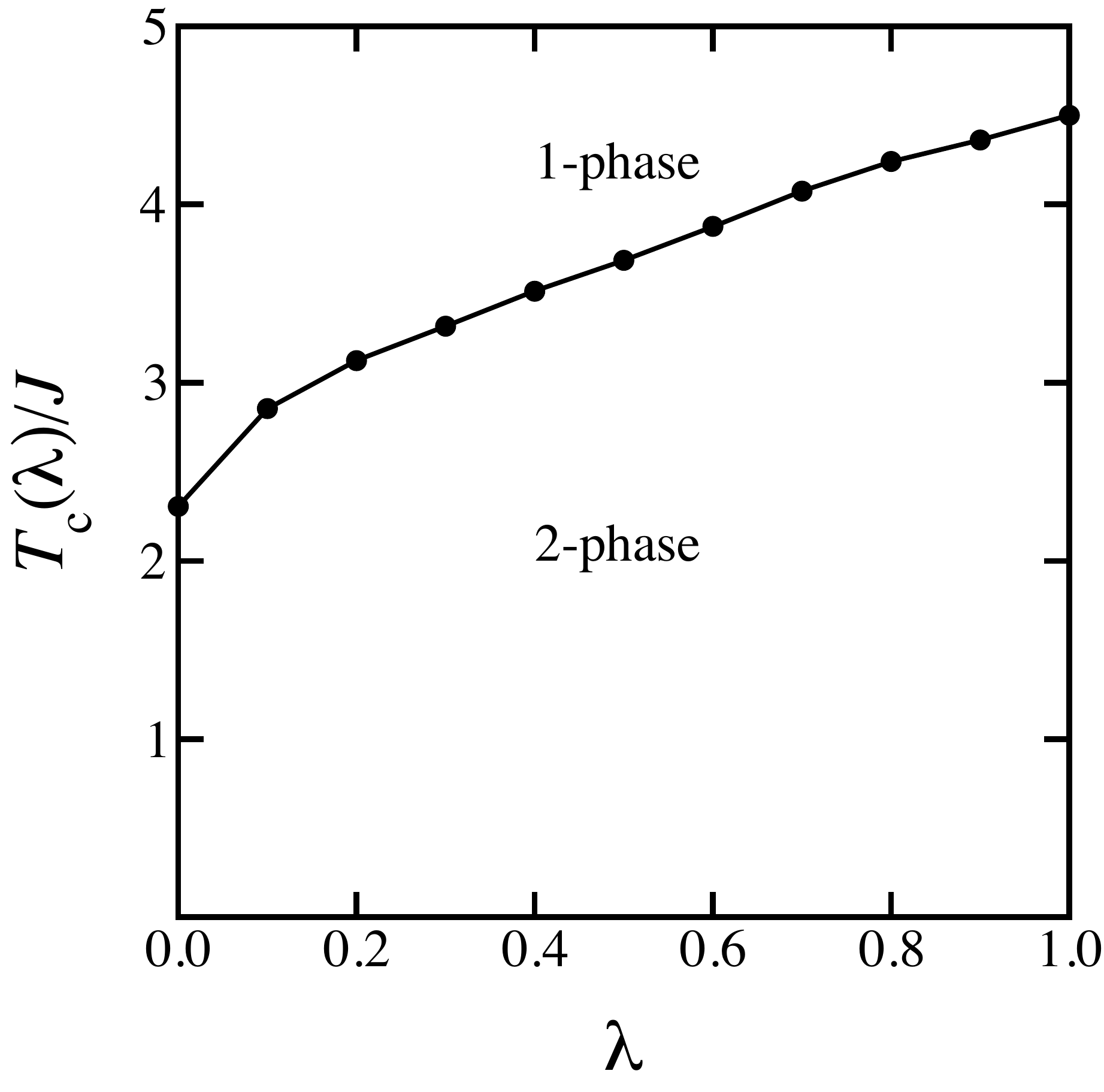}
\end{center}
\caption{\textsf{
The phase-separation temperature, $T_{\rm c}(\lambda)/J$, at the critical composition,
as a function of the interaction
parameter $\lambda$ for symmetric A/B mixtures, $\overline{S}_{i,\boldsymbol{\rho}}=0$.
The system is in a phase-separated state below the solid line, and in a one-phase
state above the line.
}}
\label{fig6}
\end{figure}

For each system size, we associate the peak position of the
specific heat with the apparent critical temperature, $T_{\rm c}(L,\lambda)$, for a system
with a finite size, $L$.
Finite-size scaling analysis is then performed in order to determine the critical temperature
for a slab of a finite $L_z$ layers in the thermodynamic limit ($L \rightarrow \infty$).
In Fig.~\ref{fig3}(b), we plot $T_{\rm c}(L,\lambda=0.1)$ as a function of $1/L$ for the
same parameters as in (a).
The plotted data are fitted with the following finite-size scaling assumption:
\begin{equation}
T_{\rm c}(L,\lambda)=T_{\rm c}(\lambda) + a L^{-1/\nu},
\label{finite}
\end{equation}
where $T_{\rm c}(\lambda)=T_{\rm c}(L \rightarrow \infty,\lambda)$ is the infinite system
critical temperature for a given $\lambda$, $a$ is a non-universal prefactor,
and $\nu$ is the 2d critical exponent for the correlation length in the $xy$-plane.
We set $\nu=1$ in our analysis, following the work by Pham Phu \textit{et al.}~\cite{Diep09},
who performed extensive MC simulations on magnetic Ising films (with $\lambda=1$)~\cite{CF76}.
We choose this 2d critical exponent for the fitting because
it was shown~\cite{Diep09} that the 2d character of the film is dominant even for $L_z=13$.
The extrapolated critical temperature for $L \rightarrow \infty$  obtained
from Fig.~\ref{fig3}(b) is $T_{\rm c}(\lambda=0.1)/J =2.85$.
We repeat this procedure for different values of the inter-layer interaction
parameter in the range of $0 \le \lambda \le 1$, and determine the
corresponding critical temperature, $T_{\rm c}(\lambda)$.
We note that the value $\nu=1$ provides a good fitting for all the
$\lambda$ values examined.

\begin{figure}[tbh]
\begin{center}
\includegraphics[scale=0.38]{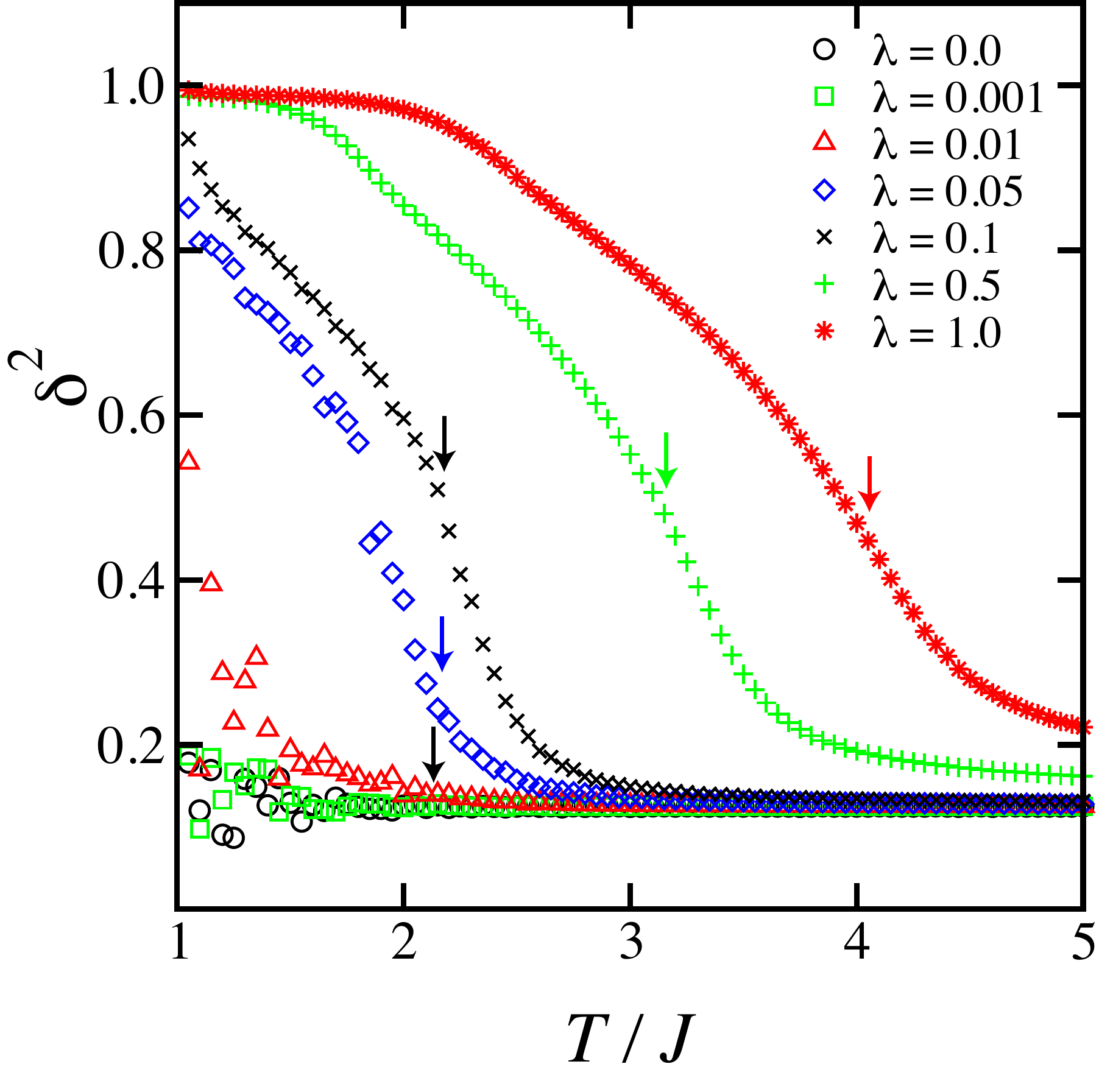}
\end{center}
\caption{\textsf{
The out-of-plane domain connectivity, $\delta^2$, defined in Eq.~(\ref{connectivity}),
as a function of the dimensionless temperature $T/J$, for different values of
$\lambda=0$, $0.001$, $0.01$, $0.05$, $0.1$, $0.5$, $1.0$.
The other parameters are $\overline{S}_{i,\boldsymbol{\rho}}=0$, $L=16$ and
$L_z=8$.
{The transition temperatures for different $\lambda$ values are indicated
by arrows.}
The value of $\delta^2$ becomes larger when domains are correlated along the $z$-direction
between different layers.
This increase in $\delta^2$ is observed for lower temperatures and larger
$\lambda$.
}}
\label{fig7}
\end{figure}

Somewhat surprisingly, finite-size effects in the $z$-direction are much weaker
as compared to those in the lateral direction.
This is shown in Fig.~\ref{fig4}, where we plot $c$ as a function of $T/J$
{when (a) $\lambda=0.1$, (b) $\lambda=0.5$ and (c) $\lambda=1$}
for different number of layers, $L_z= 4, 8, 12, 16$, while the lateral size $L=48$
is kept fixed.
{For all $\lambda$ values studied here ($0.1 \le \lambda \le 1$),
the observed peak position: $T/J \approx 2.65$ in (a), 3.30 in (b) and 4.10 in (c), is almost
independent of $L_z$, at least for $L_z \ge 8$.}
This means that, in our model with a fixed imposed magnetization (A/B composition) in
each layer, the correlation in the $z$-direction is very strong due to the cooperative
behavior of domains in different layers.

For fully equilibrated configurations, as shown in Fig.~\ref{fig2} after
$10^7$ MCS, the domains are highly connected vertically along the $z$-direction,
from the bottom layer to the top one.
This is also shown in Fig.~\ref{fig5} in which the columnar structure
of domains in different layers is clearly shown.
Hence, the correlation length in this direction exceeds $L_z$,
and the constraint of fixed magnetization (A/B composition) in each layer
{induces a strong structural correlation in the $z$-direction
even though the inter-layer interaction $J'$ is smaller than the
intra-layer interaction $J$ ($\lambda \le 1$).}
A more quantitative argument for the domain connectivity will be given later.
Because the number of layers, $L_z$, barely affects the MC results as shown in
Fig.~\ref{fig4} for $\lambda=0.1$, $0.5$ and $1$, most of the simulations were
done using $L_z=8$, which is sufficiently large in our case to observe the
asymptotic behavior of $L_z \rightarrow \infty$.
{For the anisotropic 3d Ising model without any constraint of
conserved magnetization, as previously studied in Ref.~\cite{lee01}, a very
weak system-size dependence of the apparent critical temperature was 
observed by measuring the \textit{planar} susceptibility.}

The results of finite-size scaling analysis are shown in Fig.~\ref{fig6}, where we
plot $T_{\rm c}$ as a function of $\lambda$.
The critical temperature interpolates between the 2d and 3d Ising results,
$T_{\rm c}^{\rm 2d}<T_{\rm c}(\lambda)<T_{\rm c}^{\rm 3d}$;
the exact value in 2d (corresponding to $\lambda=0$) is known to be
$T_{\rm c}^{\rm 2d}/J=2/\ln (1+\sqrt{2}) \approx 2.269$
{for square lattices}~\cite{Onsager44},
and the numerical estimate in 3d (corresponding to $\lambda=1$) is
$T_{\rm c}^{\rm 3d}/J \approx 4.511$
{for cubic lattices}~\cite{Binder01}.
These two limits are recovered in our simulations and are seen in Fig.~\ref{fig6} for
$\lambda=0$ and $1$, respectively.
Although a more detailed $\lambda$-dependent scaling behavior of $T_{\rm c}(\lambda)$
was previously discussed in the limit of very small $\lambda$~\cite{lee01,liu72}, we shall
generalize the argument for the anisotropic case of finite $\lambda$,
$0 \le \lambda \le 1$.
When $T<T_{\rm c}(\lambda)$, the stack undergoes a phase separation,
and the inplane domains rich in lipid A (spin up)
are interconnected along the $z$-direction, bridging between adjacent layers and
forming large connected domains of the same average composition.
The same feature also occurs for the B-rich domains.
Such a behavior can be clearly observed in Fig.~\ref{fig5}.

\begin{figure}[tbh]
\begin{center}
\includegraphics[scale=0.28]{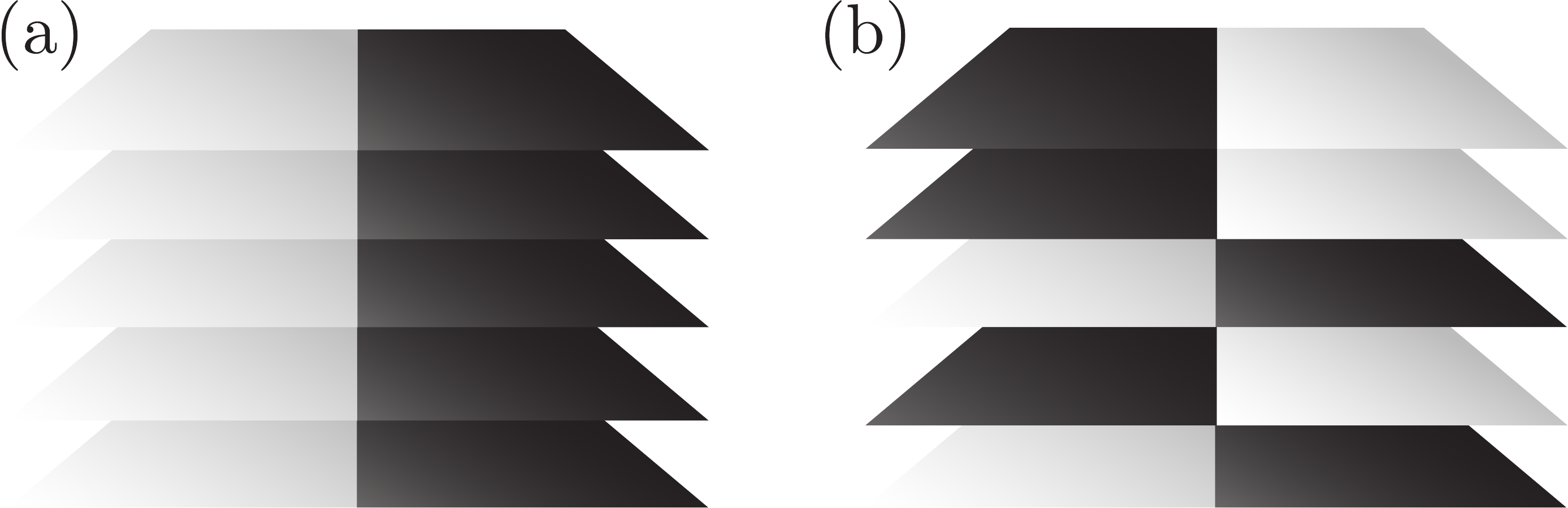}
\end{center}
\caption{\textsf{
Schematic representation of phase separated domains in a stack of membranes.
Black and white domains are rich in A and B lipids, respectively.
Two extreme cases are shown;
(a) domains are fully connected in the $z$-direction,
(b) domains are arranged at random and are disconnected.
}}
\label{fig8}
\end{figure}

In order to monitor quantitatively the degree of inter-connectivity of domains in
different layers, we define the following quantity:
\begin{equation}
\delta^2 = \frac{1}{L^2} \left\langle \sum_{\boldsymbol{\rho}}
\left( \frac{1}{L_z} \sum_i S_{i,\boldsymbol{\rho}}
- \overline{S}_{i,\boldsymbol{\rho}} \right)^2 \right\rangle,
\label{connectivity}
\end{equation}
where the average is taken over equilibrated MC configurations as explained above.
This quantity can be cast also as:
\begin{equation}
\delta^2=\frac{1}{L^2L_z^2}\sum_{\boldsymbol{\rho}}\sum_{i,j} \left\langle (S_{i,\boldsymbol{\rho}}
- \overline{S}_{i,\boldsymbol{\rho}}) (S_{j,\boldsymbol{\rho}}
- \overline{S}_{j,\boldsymbol{\rho}})\right \rangle,
\label{connectivity1}
\end{equation}
and represents a special ``magnetic susceptibility", where the correlations are taken only
along the $z$-direction and  then averaged laterally in each of the planes.
When the domains are connected along the $z$-direction, the summation over different
$i$-layers will produce a large value of $\delta$, while $\delta$ is small if the domains
are uncorrelated across the layers even for $T< T_{\rm c}(\lambda)$.
In Fig.~\ref{fig7}, we plot $\delta^2$ as a function of $T/J$ for different values
of $\lambda$, while fixing $L=16$ and $L_z=8$.
Notice that even for $\lambda$ as small as 0.05 (blue diamonds),  $\delta^2$ tends to
increase as the temperature decreases below $T_{\rm c}(\lambda)$.
This means that the domains are connected in the $z$-direction once the phase
separation takes place.
On the other hand, domains are independent and uncorrelated only when the
inter-layer interaction is extremely small, i.e, $\lambda \le 0.001$ in Fig.~\ref{fig7}.
The situation is found to be marginal when $\lambda=0.01$ (red triangles)
because $\delta^2$ then slightly deviates from zero at low temperatures.

Based on our MC results, we {conclude} that in the thermodynamic limit,
$L\to\infty$, domains will always be connected for any finite
inter-layer interaction, $J'>0$.
We give now a simple argument supporting this {conclusion}, and show that
in the limit $L \rightarrow \infty$ but with a finite number of layers, $L_z$, the domains
in different layers are uncorrelated only when $J'=0$ ($\lambda=0$) is strictly obeyed.
For the symmetric A/B case ($\overline{S}_{i,\boldsymbol{\rho}}=0$), each layer will eventually
phase separate into two semi-infinite domains: one composed by the A lipid (spin up) and the other
by the B lipid (spin down), as shown schematically in Fig.~\ref{fig8}.
When the domains are fully correlated in the $z$-direction, as in Fig.~\ref{fig8}(a),
the total free energy of the stack consists of the contributions:
\begin{equation}
F_{\rm con}=-J' L_z L^2 +F_{\rm intra},
\label{connected}
\end{equation}
where $F_{\rm intra}$ accounts for the intra-layer interactions.
On the other hand, when the inplane domains are completely random and disconnected,
as sketched in Fig.~\ref{fig8}(b), the total free energy is dominated by an entropy
contribution of arranging a random stack of A and B domains along  the $z$-direction,
\begin{equation}
F_{\rm dis}= -T L_z \ln 2 +F_{\rm intra},
\label{disconnected}
\end{equation}
with the same $F_{\rm intra}$ as before because this term is common for both
free energies.
By comparing Eqs.~(\ref{connected}) and (\ref{disconnected}), the threshold
inter-layer interaction, $(J')^{\ast}$, separating the two states, is given by:
\begin{equation}
(J')^{\ast}= \frac{T \ln2}{L^2}.
\label{transition}
\end{equation}
Notice that $(J')^{\ast}$ depends on $L$ but not on $L_z$. For finite temperatures,
it vanishes in the thermodynamic limit of $L \rightarrow \infty$.
Hence, this simple scaling argument suggests that domains are always connected in the
$z$-direction for any finite value of $J'$.
Therefore, for all $\lambda>0$, in the phase-separated region (below the critical temperature)
presented in Fig.~\ref{fig6}, domains should always form interconnected structures along
the $z$-direction.
{As shown in Eqs.~(\ref{connected}) and (\ref{disconnected}), the internal
energy scales with $L^2$, while the entropy due to the random stacking of domains does
not depend on $L$.
Hence, the entropic effect can never overcome the internal energy in the thermodynamic
limit, and leads to the stability of the columnar structure.}
This conclusion is not in agreement to that of Tayebi \textit{et al.}~\cite{Tayebi13},
who claimed that there is a ``multi-phase" state in which domains are not
aligned {and have different compositions} even in thermodynamical equilibrium.

In the simulations, $(J')^{\ast}$ can be finite due to finite-size effects.
For instance, if the temperature is chosen to be $T/J=1$ in Fig.~\ref{fig7},
the threshold value for $L=16$ can be estimated as
$\lambda^{\ast}=(J')^{\ast}/J \approx 2.7 \times 10^{-3}$.
Since $\lambda=10^{-2}$ (red triangles in Fig.~\ref{fig7}) exceeds this threshold,
the corresponding $\delta^2$ takes larger values at low temperatures.
Moreover, the very weak finite-size effects along the $z$-direction is consistent
with the lack of $L_z$-dependence of $(J')^{\ast}$ in Eq.~(\ref{transition}).

\section{Dynamics of phase separation}
\label{sec:dynamics}

\begin{figure}[tbh]
\begin{center}
\includegraphics[scale=0.38]{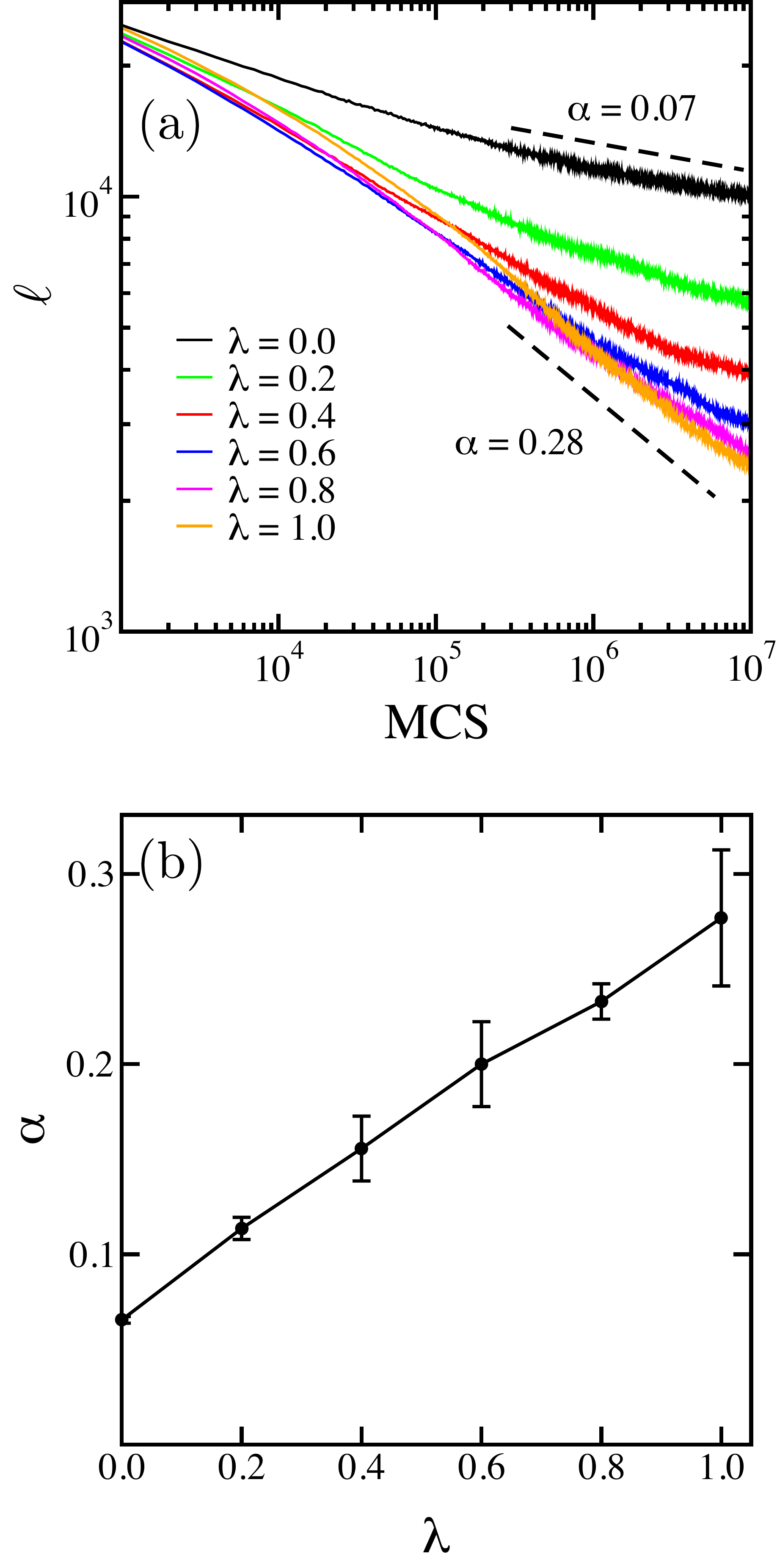}
\end{center}
\caption{\textsf{
(a) The temporal evolution of the total interface length $\ell$ as a function of
time (MCS) for different values of $\lambda=0$, $0.2$, $0.4$, $0.6$, $0.8$, $1.0$,
and for a temperature quench from the one-phase state ($T\to \infty$) into the
two-phase state at $T/J=2.0$.
The A/B mixture is symmetric, $\overline{S}_{i,\boldsymbol{\rho}}=0$,
$L=256$ and $L_z=8$.
The average over three independent MC runs is taken for each $\lambda$ value.
The two dashed lines represent a power-law behavior with exponent $\alpha=0.07$
and $0.28$, which roughly bound the two limiting behaviors of the $\lambda$-dependent
exponent, $\alpha$.
(b) The domain growth exponent $\alpha$ as a function of $\lambda$, as obtained from (a).
}}
\label{fig9}
\end{figure}

\begin{figure}[tbh]
\begin{center}
\includegraphics[scale=0.38]{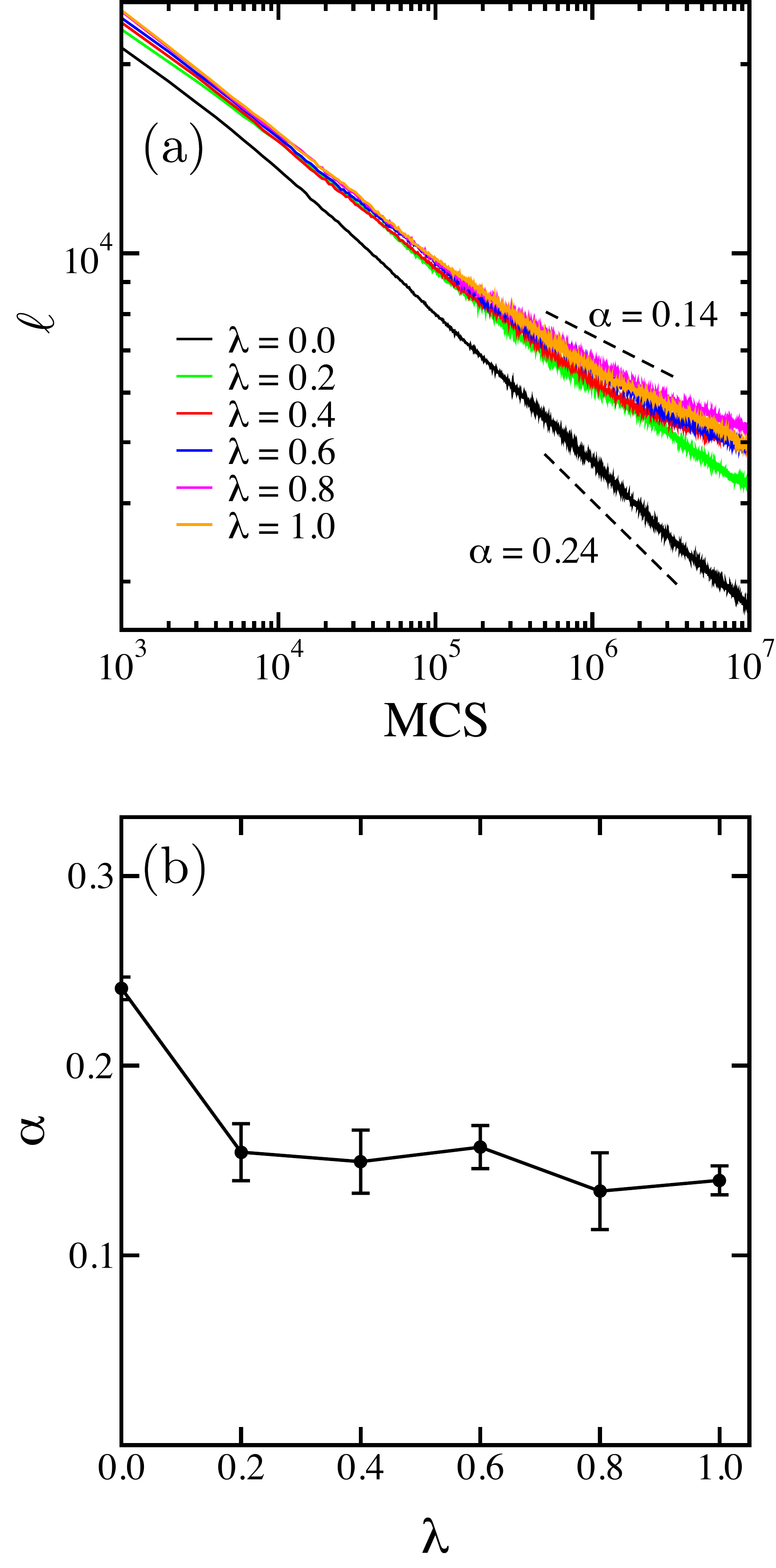}
\end{center}
\caption{\textsf{
The temporal evolution of the total interface length $\ell$ as a function of time (MCS)
for different values of $\lambda=0$, $0.2$, $0.4$, $0.6$, $0.8$, $1.0$, and for a temperature
quench from the one-phase state into the two-phase one, with final temperature
satisfying $T=0.6 T_{\rm c}(\lambda)$.
The A/B mixture is symmetric, $\overline{S}_{i,\boldsymbol{\rho}}=0$,
$L=256$ and $L_z=8$.
The average over three independent MC runs is taken for each $\lambda$ value.
The two dashed lines represent a power-law behavior with exponents $\alpha=0.14$
and $\alpha=0.24$.
(b) The domain growth exponent $\alpha$ as a function of $\lambda$, as obtained from (a).
}}
\label{fig10}
\end{figure}

We address now the effects of inter-layer interaction on the dynamics of
phase separation as the system converges towards its thermal equilibrium state.
{Under the assumption that scaling laws can be applied,}
the average domain size $R$ increases according to a temporal power-law:
$R(t) \sim t^{\alpha}$~\cite{SK_DA_Review}.
For 2d systems for which the total domain area is conserved, the average domain
size $R$ is inversely proportional to the total interface length $\ell$, i.e.,
$R \sim \ell^{-1}$~\cite{Laradji,Ramachandran10}.
This can easily be seen  because $R$ and $\ell$ are related by $\ell = 2\pi n R\sim nR$, where
$n$ is the number of domains, and the total area of all domains, $A = \pi n R^2\sim nR^2$, is a
conserved quantity.
Hence, within the scaling hypothesis, the total interface length (in 2d) should behave as
\begin{equation}
\ell(t) \sim t^{-\alpha}.
\label{scaling}
\end{equation}
In our stacked Ising model, we calculate the interface length in each of the
layers and average it over different layers.
Note that the total interface length is proportional
to the first term of the Hamiltonian in Eq.~(\ref{hamiltonian}),
which enumerates the number of bonds across the inplane A/B interface.

In Fig.~\ref{fig9}(a), we plot the temporal evolution of the
total interface length in 2d, $\ell(t)$,
(and averaged along the $z$-direction), as a function of time measured in
MC steps. The temperature quench into the two-phase region is done for a fixed temperature, $T/J=2.0<T_{\rm c}(\lambda)$, in order to mimic the experiment that is conducted at fixed room temperature.
Several values of $\lambda$ are studied, and
the other parameters are $L=256$ and $L_z=8$, with averages taken over
three independent MC runs.
For each $\lambda$ value, the scaling behavior of Eq.~(\ref{scaling}) is analyzed,
and we extract the growth exponent $\alpha$ from the late stage kinetics.
We find that for $\lambda=0$ (2d case), the growth exponent has the smallest value of
$\alpha \approx 0.07$, while for $\lambda>0$, it is a function of $\lambda$ and increases up to $\alpha \approx 0.28$,
as shown in Fig.~\ref{fig9}(b).

Although this result may explain the fact that the phase separation has an accelerated
dynamics in stacked membranes as compared to GUVs (isolated single membranes), we should
keep in mind that $T_{\rm c}(\lambda)$ increases as function of the
inter-layer coupling $\lambda>0$, as shown in Fig.~\ref{fig6}.
As long as the final quench temperature is fixed to $T/J=2.0$, the temperature quench
depth defined by $\Delta T=T_{\rm c}(\lambda) -T$ becomes larger as the value of
$\lambda$ is increased.
This may explain why the growth exponent $\alpha$ becomes larger with increasing
$\lambda$, for a fixed $T$-quench.

In order to have a better comparison between different $\lambda$ values, we
evaluate in Fig.~\ref{fig10} the growth exponent in a different way.
We now keep a constant quench ratio $T/T_{\rm c}(\lambda)=0.6$, where
$T$ is the final quench temperature, and the critical temperature $T_{\rm c}(\lambda)$
depends on $\lambda$, as shown in Fig.~\ref{fig6}.
For these deeper temperature quenches (farther from $T_c(\lambda)$),
the estimated growth exponent is $\alpha \approx 0.24$ for $\lambda=0$
(pure 2d case), and  $0.13\le \alpha \le 0.16$ for $0.2 \le \lambda\le 1.0$.
Note that the $\alpha$-values are only weakly dependent on $\lambda>0$.

Finally, we elaborate on the decreasing $\lambda$-dependence of the growth exponent
$\alpha$, and show that this behavior is consistent with the change in the
dimensionality of the stack from 2d to 3d.
In general, the growth exponent associated with phase separation depends on the
dimensionality~\cite{Bray}.
In this context, we mention the scaling argument of Binder and Stauffer on phase-separation
dynamics of particles that undergo cluster reaction and diffusion processes~\cite{Binder74}.
{Under the assumption that most particles that leave a cluster reimpinge 
on the same cluster at
later times}, the diffusion coefficient $D$ of a cluster of size $R$
was shown to scale as $D \sim R^{-(1+d)}$, where $d$ is the embedded space dimension.
If we further assume that the domain size $R$ is the only length scale in the
system, the scaling relation for a simple diffusion process is given by $R^2 \sim D t$.
This argument yields the growth exponent to be $\alpha = 1/(3+d)$.
Hence, the predicted values from this scaling conjecture are $\alpha=1/5$
for $d=2$ and $\alpha=1/6$ for $d=3$.

Our simulation results, namely, $\alpha \approx 0.24$ for $\lambda=0$ and
$\alpha \approx 0.14$ for $\lambda \ge 0.2$ compare favorably with this prediction.
The growth exponent decreases for finite $\lambda$ because the system crosses-over
from 2d to 3d.
This is due to the fact that the growing phase-separated domains are inter-connected
along the $z$-direction for $\lambda>0$.
It should be noted, however, that the absolute value of $\alpha$ obtained from the
simulation is not universal but strongly depends on the quench depth as shown in
Fig.~\ref{fig9}.
This explains why the above exponents are not in complete agreement with the
simple scaling argument of Binder and Stauffer.

\section{Concluding Remarks}
\label{sec:summary}

Motivated by recent works of Tayebi \textit{et al.}~\cite{Tayebi12,Tayebi13},
who studied experimentally and theoretically the phase separation
in stacks of multi-component lipid bilayers, we have investigated the stacked 2d Ising model
given in Eq.~(\ref{hamiltonian}).
We use a Monte Carlo simulation scheme with Kawasaki exchange dynamics that conserves
the order parameter {in each layer}, in order to investigate both equilibrium
and dynamical features.
Performing finite-size scaling analysis only in the lateral direction, while keeping the stack
thickness fixed (mimicking the experiment),
we determine the phase-transition temperature, $T_{\rm c}(\lambda)$, by
changing the inter-layer interaction parameter $\lambda=J'/J$.
As shown in Fig.~\ref{fig6}, the phase-transition temperature interpolates between
that of the 2d and 3d Ising model.

One of our main conclusions is that domains in each one of the layers are always interconnected
along the $z$-direction, forming a continuous columnar structure for any
finite inter-layer interaction $J'>0$, as shown in Fig.~\ref{fig5}.
This domain structure is in accord with the experimental findings for stacks of few dozen
to few hundred layers~\cite{Tayebi12}.
{However, the ``multi-phase" region in which there are unaligned inplane
domains with different composition, as was predicted in Ref.~\cite{Tayebi13}, is not
found in our study at thermal equilibrium.
Of course that such a ``multi-phase" state can be transiently observed before the system 
reaches its fully equilibrated state, as can be observed in Figs.~\ref{fig2} and \ref{fig5}.}

We have also investigated the temporal evolution of domain formation in the stacked
2d Ising model.
When the inter-layer interaction $\lambda$ increases, the phase separation appears
to have an  accelerated dynamics as can be seen by the larger values of the growth
exponent, $\alpha$, shown in Fig.~\ref{fig9}(b).
However, these larger $\alpha$ values are mainly due to an increase in the
phase-transition temperature, $T_{\rm c}(\lambda)$, as function of $\lambda$;
thus, a larger effective temperature quench, $\Delta T=T_{\rm c}(\lambda)-T$,
for fixed $T$.
When the final temperature quench $T$ is fixed relative to the phase-transition temperature
as shown in Fig.~\ref{fig10} for $T=0.6 T_{\rm c}(\lambda)$, the growth exponent even
decreases
{as the $\lambda$ value is increased.}
Our numerical findings for the growth exponent $\alpha$ are different than the value of
$\alpha \approx 0.455$, as found in the experiment~\cite{Tayebi12}.
One possible explanation for this discrepancy can be the lack of hydrodynamic interactions
in our MC simulations~\cite{Stanich}.

In this work, we have mainly discussed the case of $\overline{S}_{i,\boldsymbol{\rho}}=0$,
corresponding to the critical composition of the A/B lipid mixture.
{Currently, we are investigating the dynamics of phase separation for
off-critical compositions, $\overline{S}_{i,\boldsymbol{\rho}} \neq 0$ [see
Fig.~\ref{fig5}(b)]. For such compositions, the phase-transition temperature is smaller than the critical
temperature.}
In the present simulations, the average A/B lipid composition (order parameter of the
Ising model) in each bilayer is restricted to stay the same.
In the future, we plan to study membrane stacks where each {layer} has
a different but fixed composition~\cite{Sornbundit13}.
Furthermore, since it is known from simulations
that the presence of a supporting solid substrate affects the dynamics of membrane
domain growth~\cite{Ngamsaad11}, it will be of interest to incorporate this substrate
effect in future studies.

\begin{acknowledgments}

We thank H.\ T.\ Diep, J.-B.\ Fournier, T.\ Kato, R.\ Okamoto, H.\ Orland,
P. Sens, S.\ Shimobayashi, M. Turner for useful discussions.
T.H. acknowledges support of Tokyo Metropolitan University for an international
student exchange program and the hospitality of Tel Aviv University.
S.K. acknowledges support from the Grant-in-Aid for Scientific Research on
Innovative Areas ``Fluctuation and Structure" (Grant No.\ 25103010) from the Ministry
of Education, Culture, Sports, Science, and Technology of Japan,
the Grant-in-Aid for Scientific Research (C) (Grant No.\ 24540439)
from the Japan Society for the Promotion of Science (JSPS),
and the JSPS Core-to-Core Program ``International Research Network
for Non-equilibrium Dynamics of Soft Matter".
D.A. acknowledges support from the Israel Science Foundation under Grant
No.\ 438/12 and the United States--Israel Binational Science Foundation under
Grant No.\ 2012/060.
\end{acknowledgments}


\end{document}